\documentclass[journal]{IEEEtran}
\IEEEoverridecommandlockouts

\usepackage{amsmath}

\usepackage{enumitem}
\usepackage{graphicx}
\usepackage{color}
\usepackage{amsmath}
\usepackage{mathtools}
\usepackage{multicol}
\usepackage{multirow}
\usepackage[english]{babel}
\usepackage{blindtext}
\usepackage{algorithm}
\usepackage{algorithmic}
\usepackage{balance}
\usepackage{amsfonts}
\usepackage{bm}
\usepackage{stfloats}
\usepackage{subfig}
\usepackage{amsthm}
\usepackage{amssymb}
\usepackage{setspace}
\usepackage[nosort]{cite}
\usepackage{CJK}
\usepackage{cite}
\usepackage{caption}
\usepackage{comment}
\usepackage{array,multirow}
\usepackage{graphicx}

\theoremstyle{plain}

\usepackage[table]{xcolor}

\begin{document}

\captionsetup[figure]{labelformat={default},labelsep=period,name={Fig.}}

\title{Modality-Decoupled Federated Learning for Privacy-Preserving Embodied Intelligence in 6G}

\author{Zhuodong Liu,
        Xiangyu Li,
        Chunhong Yuan,
        Hongyang Du,~\IEEEmembership{Member,~IEEE},
        Bodong Shang,~\IEEEmembership{Member,~IEEE},
        Qingqing Wu,~\IEEEmembership{Senior Member,~IEEE},
        Tony Q. S. Quek,~\IEEEmembership{Fellow,~IEEE},
        and Mohsen Guizani,~\IEEEmembership{Fellow,~IEEE}
\thanks{Zhuodong Liu is with the Artificial Intelligence Thrust, The Hong Kong University of Science and Technology (Guangzhou), Guangzhou 511453, China.}
\thanks{Xiangyu Li (corresponding author) is with the Department of Electronic Engineering, Shanghai Jiao Tong University, Shanghai 200240, China, and also with the School of Electronic Science and Technology, Eastern Institute of Technology, Ningbo, Zhejiang 315200, China (e-mail: xiangyuli@sjtu.edu.cn).}
\thanks{Chunhong Yuan is with the Faculty of Control Systems and Robotics, ITMO University, St. Petersburg, 197101, Russia.}
\thanks{Hongyang Du is with the Department of Electrical and Electronic Engineering, University of Hong Kong, Hong Kong SAR, China.}
\thanks{Bodong Shang is with the School of Electronic Science and Technology, Eastern Institute of Technology, Ningbo, Zhejiang 315200, China.}
\thanks{Qingqing Wu is with the Department of Electronic Engineering, Shanghai Jiao Tong University, Shanghai 200240, China.}
\thanks{Tony Q. S. Quek is with the Information Systems Technology and Design Pillar, Singapore University of Technology and Design, Singapore 487372.}
\thanks{Mohsen Guizani is with the Machine Learning Department, Mohamed Bin Zayed University of Artificial Intelligence, Abu Dhabi, United Arab Emirates.}
}

\maketitle

\begin{abstract}
Sixth-generation (6G) wireless networks are expected to provide a key infrastructure for large-scale embodied intelligence, where heterogeneous robots collaborate through low-latency connectivity, edge intelligence, and distributed sensing. Vision-language-action (VLA) models offer a foundation by integrating visual perception, language understanding, and action generation into a unified closed-loop policy. 
However, training and adapting VLA models to distributed robotic agents introduce challenges in privacy protection, communication efficiency, and model heterogeneity. Existing federated learning (FL) methods overlook the intrinsic differences among vision, language, and action pathways in parameter scale, privacy exposure, update dynamics, and tolerance to compression or perturbation. 
To address this issue, this article proposes FedMVLA, a modality-decoupled FL framework for privacy-preserving embodied intelligence in 6G networks. FedMVLA incorporates three mechanisms: modality-aware federated aggregation (MAFA), modality-aware privacy allocation (MAPA), and modality-aware communication compression (MACO), together with a modality-sliced transport design that routes the precision-critical action stream through a protected ultra-reliable low-latency slice.
A case study on federated robotic manipulation over the Third Generation Partnership Project (3GPP)-based wireless substrate, covering fading, co-channel interference, and malicious jamming, shows that FedMVLA achieves an 84.8\% task success rate, exceeds FedAvg by 22.2 percentage points, sustains a widening margin when scaling to 128 clients across eight cells, and reduces the schedule-averaged per-client uplink model-update payload by 95.6\% (approximately 96\%), while keeping the 95th percentile (p95) of the round-critical uplink completion time near 1.5\,s. These results indicate that modality-decoupled FL provides a pathway toward scalable, trustworthy, and 6G-native embodied intelligence, and open issues and directions are highlighted for future research.
\end{abstract}

\IEEEpeerreviewmaketitle

\section{Introduction}
With massive machine-type communication (mMTC), ultra-reliable low-latency communication (URLLC), integrated sensing and communication, and artificial intelligence (AI)-native edge intelligence, sixth-generation (6G) wireless networks are envisioned as a fundamental infrastructure for large-scale embodied intelligence \cite{bariah2024ai}. They can connect heterogeneous robots to edge intelligence while supporting real-time perception, low-latency coordination, and collaborative decision-making across distributed physical environments. As robotic systems move from isolated operation to collaborative deployment, embodied agents must continuously adapt to different tasks, environments, and user requirements using locally collected interaction data. Such data are distributed, bandwidth-intensive, and privacy-sensitive, making centralized collection and retraining increasingly impractical.

\begin{figure*}[tp]
\begin{center}
\setlength{\abovecaptionskip}{+0.2cm}
\setlength{\belowcaptionskip}{-0.0cm}
\centering
  \includegraphics[width=5.6in]{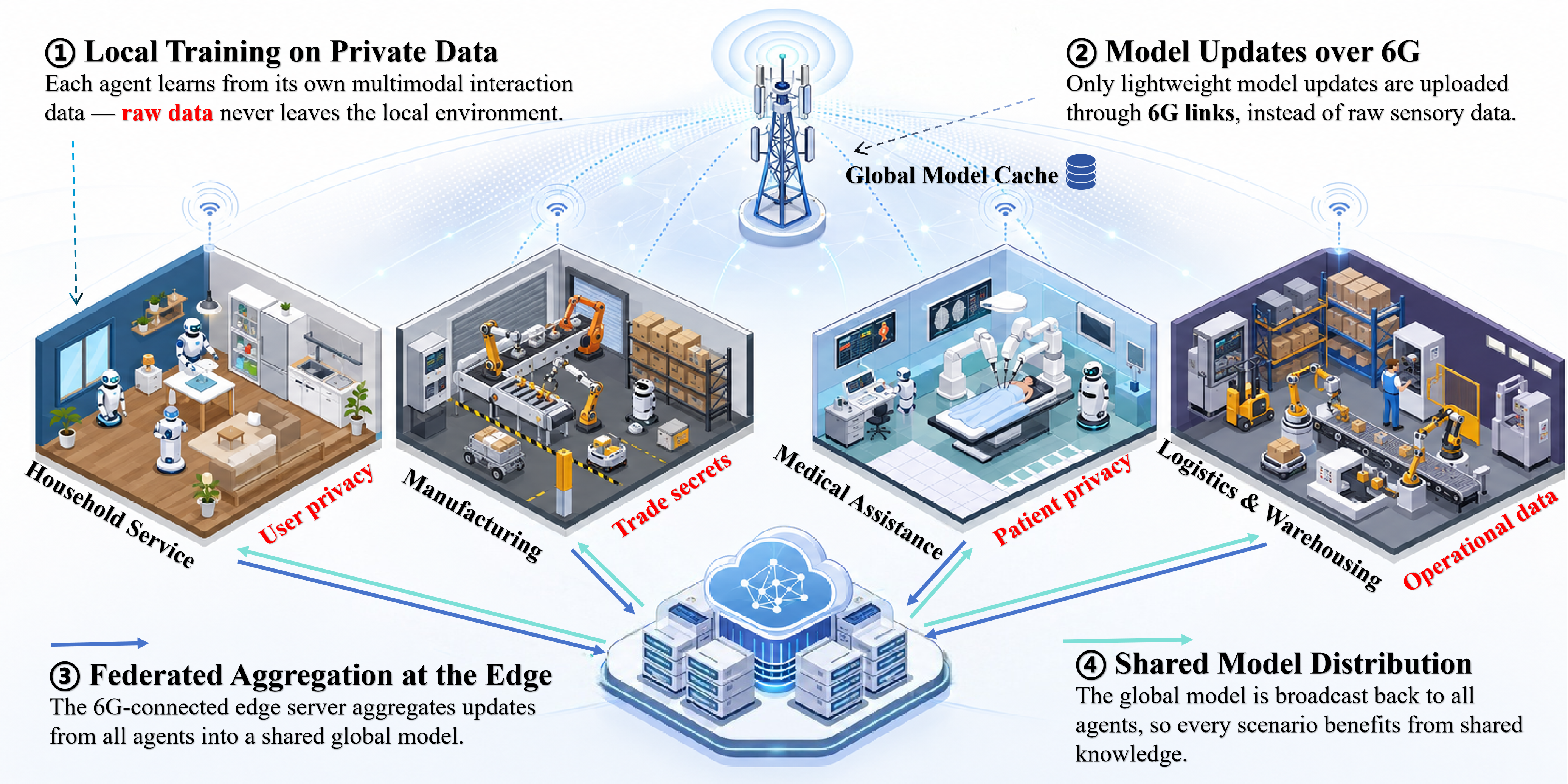}
\renewcommand\figurename{FIGURE}
\caption{\scriptsize Representative application scenarios of federated embodied intelligence in 6G networks. Heterogeneous robotic agents, including industrial manipulators, medical assistants, logistics robots, and household service robots, collaboratively train shared VLA models through 6G-connected edge servers without exchanging raw interaction data.}
\label{fig_scenario}
\end{center}
\vspace{-6mm}
\end{figure*}

Vision-language-action (VLA) models provide a promising foundation for this setting by integrating visual observations, natural-language instructions, and physical actions into a unified ``perceive-understand-act'' policy \cite{kim2024openvla}. A typical VLA model comprises a vision encoder for extracting scene representations, a language backbone for instruction understanding and semantic reasoning, and an action decoder for generating executable control sequences. Leveraging large-scale vision-language representations, VLA policies can recognize objects, reason about spatial relations, interpret task semantics, and translate these capabilities into physical behavior. This unified perception--reasoning--action structure is well suited to collaborative robotic systems operating over distributed 6G edge networks, as illustrated in Fig.~\ref{fig_scenario}.

The same multimodal structure, however, complicates distributed adaptation. Scaling VLA policies from laboratory or simulation environments to heterogeneous robotic deployments requires diverse demonstrations collected across embodiments, scenes, and users. Visual observations can reveal spatial layouts and identities; language instructions can expose intentions and routines; and action trajectories can reflect physical behavior or health-related patterns. Leakage or manipulation of either the raw data or the resulting model updates can threaten privacy, task integrity, and physical safety.

\captionsetup{font={scriptsize}}

Federated learning (FL) provides a natural paradigm for addressing these challenges. 
In FL, each client trains a model locally on private data and uploads only model updates to a server for aggregation, without directly sharing raw data \cite{mcmahan2017communication}. 
This mechanism aligns well with the distributed edge architecture of 6G networks by keeping embodied interaction data local while reducing the need to transmit large-scale multimodal datasets. 
Recent studies have applied FL to embodied intelligence and robotic systems, enabling distributed agents to collaboratively learn navigation or manipulation policies without exposing local interaction data \cite{zhou2022fedvln,miao2025fedvla}. However, directly applying conventional FL methods to VLA-driven embodied intelligence remains insufficient. 

The consequences of ignoring the internal structure of VLA models are tangible in deployment. Averaging vision encoders across visually dissimilar sites biases shared features toward dominant environments, so a household robot may mis-localize objects and mis-grasp after being aggregated with factory peers. A single differential privacy (DP) budget strong enough to protect action trajectories injects so much noise into instruction understanding that ``place the cup on the upper shelf'' is executed on the wrong shelf. Compressing action updates like any other tensor introduces millimeter-level end-effector errors that surface as manipulator jitter and failed grasps.
Existing FL studies mainly address client-level heterogeneity in data distributions, computational capabilities, and network conditions. VLA models introduce an additional modality-level dimension across vision, language, and action. The vision encoder is parameter-heavy and sensitive to camera viewpoint, illumination, and scene layout; the language pathway is commonly adapted through lightweight modules and is comparatively tolerant to quantization; and the action pathway is tightly coupled to embodiment, dynamics, and control precision. Treating these pathways uniformly can therefore create three failures: conflicting or diluted updates under a single aggregation topology; either insufficient protection or excessive utility loss under one privacy budget; and compression that wastes bandwidth on tolerant parameters while degrading precision-critical action updates.

Current federated embodied intelligence methods only partially address this internal structure. FedVLN uses partial-module aggregation for vision-language navigation \cite{zhou2022fedvln}, whereas FedVLA applies mixture-of-experts aggregation to task heterogeneity in robotic manipulation \cite{miao2025fedvla}. These studies establish the feasibility of federated embodied learning, but their primary optimization unit remains the client, task, or expert. They do not jointly coordinate modality-specific aggregation, privacy allocation, communication compression, and transport protection across all three VLA pathways. This gap motivates a modality-decoupled framework tailored to the structure of VLA policies.

Motivated by these observations, this article proposes FedMVLA, a modality-decoupled FL framework for privacy-preserving embodied intelligence in 6G networks. Its core idea is to separate local updates along the vision, language, and action boundaries and then assign each stream an appropriate federated treatment. Modality-aware federated aggregation (MAFA) selects pathway-specific aggregation topologies and synchronization frequencies according to cross-client divergence and convergence behavior. Modality-aware privacy allocation (MAPA) jointly considers privacy sensitivity and perturbation tolerance when distributing a shared differential-privacy budget. Modality-aware communication compression (MACO) matches sparsification or quantization to the precision sensitivity of each pathway. These mechanisms form a coordinated pipeline: MAFA determines the aggregation and exposure structure, MAPA calibrates client-side clipping and noise, and MACO reduces the transmitted payload without sacrificing action-update fidelity.
The main contributions of this article are summarized as follows:
\begin{itemize}
    \item \textbf{Modality-Level Heterogeneity Analysis:} 
    We identify modality-level heterogeneity as a fundamental challenge in federated VLA learning for 6G-enabled embodied intelligence, where the vision, language, and action pathways differ in parameter scale, update dynamics, privacy exposure, and compression sensitivity. 
    This shifts the focus of FL design from client- or task-level heterogeneity to the internal multimodal structure of VLA models.
    \item \textbf{Modality-Decoupled Framework Design:} 
    To address the above challenge, we propose FedMVLA, a modality-decoupled FL framework that separates VLA model updates into vision, language, and action streams and processes them through three coordinated mechanisms---MAFA, MAPA, and MACO---that jointly customize aggregation, privacy allocation, and communication compression for privacy-preserving and communication-efficient embodied intelligence over 6G edge networks.
    \item \textbf{Validation and Deployment Insights:} 
    We validate FedMVLA on the Third Generation Partnership Project (3GPP)-based wireless substrate against conventional and federated embodied baselines, showing gains in task success, privacy--utility trade-off, uplink latency, and robustness at scale. We further discuss safety-aware aggregation, open radio access network (O-RAN) orchestration, and semantic communication for future 6G deployment.
\end{itemize}

\section{Foundations of Federated Embodied Intelligence}

\subsection{Federated VLA Learning in 6G Edge Networks}

FL provides a distributed training paradigm in which each robot learns from local private data and uploads only model updates for server-side aggregation \cite{mcmahan2017communication}. This paradigm fits 6G-connected embodied systems because it keeps high-volume interaction data at the edge while allowing heterogeneous agents to benefit from shared experience. Low-latency connectivity, massive access, and AI-native edge orchestration further support collaborative learning among service robots, industrial manipulators, mobile agents, and assistive devices \cite{bariah2024ai}.

Federated embodied learning nevertheless combines several sources of difficulty. Clients differ in computational capability, sensor configuration, link quality, embodiment structure, and local data distribution, creating both system and statistical heterogeneity. Large multimodal policies also impose substantial memory and communication burdens. Moreover, retaining raw data locally does not by itself eliminate privacy risk: model updates can reveal sensitive information through attacks such as gradient inversion, motivating differential privacy, secure aggregation, and restricted update exposure \cite{du2025flguard}.

VLA models add a modular source of heterogeneity to these client-level differences. Their vision encoder, language backbone, and action decoder map observations and instructions to executable behavior \cite{kim2024openvla}, but these modules serve different functions and exhibit different parameter scales, update dynamics, privacy profiles, and tolerances to perturbation. Large cross-embodiment datasets demonstrate the value of diverse trajectories, yet collecting and adapting such data across real deployments remains expensive and privacy-sensitive. Effective federated VLA learning should therefore account explicitly for structural differences among the three pathways rather than treating the policy as an indivisible model.

\subsection{Modality-Level Heterogeneity in Federated VLA Training}
Existing FL research addresses client-level heterogeneity, such as differences in device capability, network condition, and local data distribution. 
However, VLA models introduce another critical dimension: modality-level heterogeneity. 
As summarized in Table~\ref{table_modality}, the vision encoder, language backbone, and action decoder differ significantly in parameter scale, training strategy, output representation, cross-client divergence, privacy exposure, and compression sensitivity.

\begin{table*}[t]
\centering
\caption{Heterogeneous Characteristics of Modality-Specific Modules in VLA Models (M and B denote million and billion parameters, respectively.)}
\label{table_modality}

\begingroup
\footnotesize
\setlength{\tabcolsep}{4.5pt}
\renewcommand{\arraystretch}{1.16}
\setlength{\arrayrulewidth}{0.5pt}

\begin{tabular*}{\textwidth}{@{\extracolsep{\fill}}
>{\raggedright\arraybackslash}p{3.00cm}
>{\raggedright\arraybackslash}p{3.55cm}
>{\raggedright\arraybackslash}p{4.35cm}
>{\raggedright\arraybackslash}p{4.95cm}
@{}}
\hline
\textbf{Property}
& \textbf{Vision Encoder}
& \textbf{Language Backbone}
& \textbf{Action Decoder} \\
\hline

Representative modules
& SigLIP, DINOv2
& LLaMA-2, Gemma
& Multilayer perceptron (MLP) head; diffusion policy \\

Parameter scale
& 300M--1B
& 2B--7B
& 30M--300M \\

Typical tuning strategy
& Frozen or slow fine-tuning
& Low-rank adaptation (LoRA) or partial fine-tuning
& Head or adapter fine-tuning \\

Output representation
& Continuous embeddings
& Discrete token logits
& Continuous or discretized \\

Cross-client divergence
& High (varied cameras and scenes)
& Low (standardized instructions)
& High (heterogeneous embodiments) \\

Privacy exposure risk
& Spatial layout and user identity
& Task intent and daily routines
& Physical behavior and health status \\

Compression sensitivity
& Moderate (spatial redundancy)
& Low (discrete and error-correctable)
& High (precision-critical signals) \\

\hline
\end{tabular*}

\endgroup
\vspace{-6mm}
\end{table*}

These modality-level differences directly shape federated optimization. For aggregation, visual features shift with camera viewpoint, illumination, background texture, and scene layout, while action updates also vary with robot kinematics and control dynamics. A single global topology can therefore mix incompatible updates or allow dominant environments to bias shared representations. For privacy, the three pathways expose different information and lose utility at different rates under the same perturbation strength, so a uniform budget can either under-protect sensitive updates or over-perturb fragile ones. For communication, language updates tolerate coarse quantization, visual updates contain exploitable redundancy, and action-head updates remain sensitive to precision loss.

Section~IV measures these asymmetries through matched noise and compression tests, with utility gaps spanning more than an order of magnitude. The empirical differences motivate a coordinated modality-aware design rather than three unrelated heuristics: aggregation topology addresses cross-client divergence, privacy allocation addresses the sensitivity-tolerance trade-off, and compression addresses pathway-specific precision requirements.

\subsection{Comparison with Related Methods}

FL aggregation methods have evolved from standard parameter averaging to more robust optimization under heterogeneous data distributions. 
FedAvg establishes the basic weighted aggregation paradigm \cite{mcmahan2017communication}, FedProx introduces proximal regularization to mitigate system and statistical heterogeneity \cite{li2020federated}, and SCAFFOLD corrects client drift through control variates under non-independent and identically distributed (non-IID) data \cite{karimireddy2020scaffold}. 
However, these methods generally treat the model as an indivisible whole and do not explicitly consider modality-level structural differences.

Recent studies have begun to explore FL for embodied intelligence. 
FedVLN applies FL to vision-language navigation and aggregates selected modules during the pre-exploration stage \cite{zhou2022fedvln}.
FedVLA extends FL to VLA-based robotic manipulation and proposes a dual-gating mixture-of-experts architecture with expert-driven aggregation to address task-level heterogeneity \cite{miao2025fedvla}. 
Although these works advance privacy-preserving embodied intelligence, they mainly focus on client-level or task-level heterogeneity. 
In contrast, FedMVLA treats modality-level heterogeneity as the primary optimization unit and jointly designs aggregation, privacy allocation, and communication compression for vision, language, and action pathways. 
Both are included as empirical baselines in Section~IV.

\section{Decoupling Modalities, Unifying Intelligence: The FedMVLA Framework}

\begin{figure*}[tp]
\begin{center}
\setlength{\abovecaptionskip}{+0.1cm}
\setlength{\belowcaptionskip}{-0.0cm}
\centering
  \includegraphics[width=6.4in]{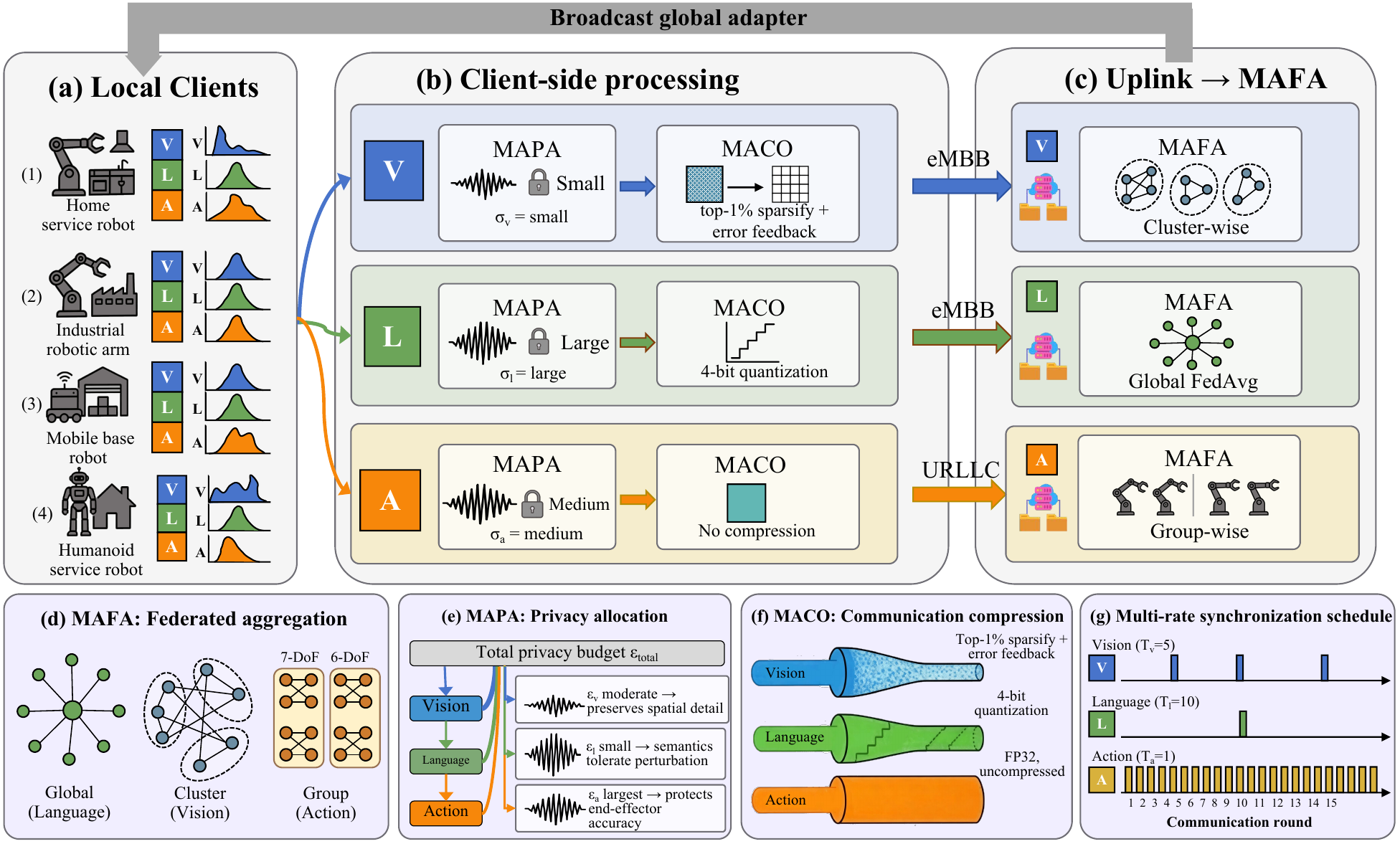}
\renewcommand\figurename{FIGURE}
\caption{\scriptsize Overview of FedMVLA. (a) Heterogeneous local clients hold non-IID vision (V), language (L), and action (A) updates. (b) Client-side processing: MAPA calibrates per-modality clipping and noise injection before MACO applies modality-matched compression, so compression and transport are post-processing with respect to the differential privacy guarantee. (c) Each stream is uplinked over its transport slice---eMBB for vision and language and URLLC for action---and MAFA aggregates vision cluster-wise, language globally, and action group-wise by embodiment, after which the global adapter is broadcast back to all clients. Panels (d)--(g) detail MAFA's aggregation topologies, MAPA's privacy allocation, MACO's per-modality compression, and the multi-rate synchronization schedule.}
\label{fig_framework}
\end{center}
\vspace{-6mm}
\end{figure*}

Building on the modality-level heterogeneity identified in Section~II-B, FedMVLA systematically decouples the federated training of VLA models along modality boundaries, applying tailored strategies across three dimensions: aggregation, privacy, and communication. As illustrated in Fig.~\ref{fig_framework}, each client performs local training and then separates its model updates into vision, language, and action streams via a modality decoupling module. 
Each stream then follows a client--server pipeline: MAPA calibrates per-modality clipping and noise injection on the client before transmission, MACO compresses each stream to match its transport slice, and the server-side MAFA aggregates each stream under its own topology and schedule; a modality reassembly module reunifies the processed updates into a consolidated adapter for redistribution to all clients. Since noise injection precedes compression, the differential privacy guarantee is preserved under post-processing.

The framework is organized around two intrinsic pathway attributes. Cross-client divergence determines how widely an update should be aggregated, whereas perturbation and compression tolerance determine how strongly it can be protected and compressed. The action pathway exhibits pronounced cross-embodiment variation and severe precision sensitivity; it therefore uses embodiment-wise aggregation, moderate privacy noise, and full-precision transmission. The language pathway is more globally aligned and comparatively tolerant to perturbation, so it uses global aggregation, stronger privacy noise, and four-bit quantization. The vision pathway occupies an intermediate regime: it benefits from knowledge sharing, but only among clients with similar visual environments, and its large update size motivates aggressive sparsification.

This design also explains why the three mechanisms are coupled rather than independent add-ons. MAFA determines which clients contribute to each pathway and how often that pathway is released; MAPA assigns noise under the resulting release pattern; and MACO selects a representation compatible with both the pathway's precision requirement and its transport slice. The reassembled adapter, therefore, reflects coordinated modality-level decisions within one federated training loop.

\captionsetup{font={scriptsize}}

\subsection{Aggregating by Modality, Not by Client (MAFA)}
MAFA assigns each modality a differentiated aggregation topology based on its cross-client divergence. 
The \textbf{action pathway} adopts group-wise aggregation restricted to same-embodiment clients to prevent gradient conflicts. 
The \textbf{vision pathway} faces high cross-client divergence due to varying cameras, lighting, and scene layouts, yet it still benefits from shared visual knowledge; it therefore applies cluster-wise aggregation, where clients are grouped by the cosine similarity of their vision-encoder gradients so that updates from visually similar environments are aggregated together.
The \textbf{language pathway} performs global aggregation given its minimal cross-client divergence. The three topologies follow a layered, multi-rate schedule: action participates every round, vision synchronizes every several rounds, and language synchronizes least frequently. 
Concretely, vision clusters are obtained by spectral clustering on an exponentially averaged cosine-similarity matrix of vision-adapter gradients, following the gradient-similarity principle of clustered FL, re-clustered every ten rounds with at most eight clusters; the schedule is set to $(T_a, T_v, T_l)=(1, 5, 10)$ rounds. Both the schedule and the three topologies are grounded empirically in Section~IV.

\subsection{Allocating Privacy Where It Matters Most (MAPA)}

The privacy profile of each modality is governed by two factors: \emph{privacy sensitivity} (action highest, vision moderate, language lowest) and \emph{perturbation tolerance} (language highest, vision moderate, action lowest). Neither factor alone determines a useful allocation. MAPA decomposes a shared budget across the three pathways, and R\'enyi differential privacy (RDP) \cite{mironov2017renyi} composes the releases at the record level, with one demonstration trajectory as the unit of protection. The language pathway receives the smallest budget and largest noise because semantic performance remains comparatively stable under perturbation. Vision receives less noise to preserve the spatial detail needed for localization and grasp planning. Action receives moderate noise because stronger perturbation directly degrades end-effector accuracy and task success.

The formal guarantee is derived from clipping, Gaussian perturbation, and RDP composition. Group-wise aggregation and secure aggregation within each embodiment group additionally restrict server-side exposure to group sums; these mechanisms complement, rather than replace, the formal privacy guarantee. This distinction is important for the action pathway, where utility limits the amount of perturbation that can be introduced.

Clipping thresholds follow quantile-based adaptive clipping \cite{andrew2021adaptive}: each modality tracks a target quantile of its gradient norms, avoiding a fixed threshold whose relative noise grows as training gradients shrink. The multi-rate schedule also reduces privacy composition because a modality contributes to the composed mechanism only once every $T_m$ rounds. Since MACO operates on already-noised updates, compression and transport remain post-processing with respect to the reported guarantee. Together, pre-transmission noise, restricted aggregation, secure aggregation, and reduced release frequency limit the exposure of precision-critical action updates while preserving their utility.

\subsection{Compressing Smart, Not Uniform (MACO)}

MACO matches each pathway with a compression scheme consistent with its measured precision tolerance. The vision pathway applies top-$k$ sparsification, retaining the largest 1\% of parameter updates with error feedback \cite{lin2018deep}; after values and indices are counted, the payload is about $1/59$ of a 32-bit floating-point (FP32) representation. The residual feedback buffer accumulates discarded components and re-injects them in later rounds, limiting the long-term optimization bias. This aggressive sparsification is effective because visual adapters contain substantial redundancy across spatial patches and background features.

The language pathway uses group-wise four-bit quantization with group size 128. Its payload is approximately one-eighth of FP32 before scale metadata, and the micro-benchmark in Section~IV shows only a small task-success loss. The action pathway remains in FP32 because quantization-induced parameter error propagates through the action decoder to end-effector motion and task success. Combined with MAFA's multi-rate schedule, these differentiated rates amortize the originally dominant vision and language traffic while preserving the per-round fidelity of the action update. MACO therefore reduces communication by exploiting measured tolerance rather than applying a uniform rate to the entire trainable model.

\vspace{-2mm}
\subsection{Interaction with 6G Network Architecture}

FedMVLA aligns with the hierarchical, adaptive, and service-differentiated operation envisioned for 6G by organizing updates according to pathway-specific communication, privacy, computation, and reliability requirements. Unlike a datacenter fabric, the wireless medium introduces time-varying bandwidth, fading, interference, retransmission, deadline misses, and exposure of updates over open channels. At the same time, the three pathways impose conflicting transport demands: action requires frequent and reliable delivery, vision dominates the payload, and language can tolerate both delay and aggressive quantization. A monolithic transport pipeline cannot satisfy these requirements efficiently.

Specifically, MAFA's multi-rate, modality-specific schedule can be orchestrated through the dual-loop control of the O-RAN non-real-time RAN intelligent controller (Non-RT RIC) and near-real-time RAN intelligent controller (Near-RT RIC) via the A1 and E2 interfaces. MAPA's adaptive noise scaling can be driven by RAN-native AI that adjusts privacy budgets based on real-time channel states and client participation, whereas MACO's differentiated compression can use 6G network slicing to route action updates through a URLLC slice for reliability and vision updates through an enhanced mobile broadband (eMBB) slice for throughput. 
The slicing, scheduling, and deadline mechanisms are implemented in Section~IV, while the O-RAN control procedures are discussed as a deployment path in Section~V-B.

When link quality degrades, the server prioritizes the compact, precision-critical action update (approximately 1.2\,MB) and defers vision or language updates to their next synchronization point. The corresponding staleness is recorded and bounded by the next scheduled synchronization. This policy keeps large, delay-tolerant streams off the round-critical path without treating them as permanently lost.

The threat model distinguishes digital-link failures from value perturbation. With cyclic redundancy check (CRC) protection and hybrid automatic repeat request (HARQ), fading, co-channel interference, and jamming produce retransmissions, erasures, and staleness rather than undetected changes to gradient values; Section~IV reports residual loss and staleness. Analog over-the-air aggregation and falsified client updates, which can directly perturb values, are treated as future directions in Sections~V-C and V-A. The action stream uses a URLLC slice modeled on representative industrial-control targets in 3GPP Technical Specification (TS)~22.104, including 10\,ms end-to-end latency and at least 99.999\% communication-service availability \cite{3gpp22104}. Because an action update spans many packets, its update-level completion time is on the order of one second, motivating full-precision transmission rather than aggressive lossy compression. Vision and language share an eMBB slice with more relaxed latency tolerance, making sparsification, quantization, and deferred synchronization viable.

\section{Case Study: Validating Modality Decoupling in Federated Robotic Manipulation}

This section evaluates FedMVLA in simulated federated robotic manipulation over a 3GPP-based wireless substrate, covering convergence, scalability, robustness, privacy--utility trade-offs, and communication latency against conventional and embodied FL baselines.

\subsection{Experimental Setup}
The VLA backbone follows an OpenVLA-inspired design \cite{kim2024openvla}, comprising a SigLIP vision encoder (400M parameters, LoRA rank 16, 4.0M trainable), a LLaMA-2 language backbone (7B base, LoRA rank 8, 4.2M trainable), and a separately trained diffusion-based action head (50M parameters, of which only the conditioning and output layers, 0.3M, are trained; action chunk length 16). 
The 8.5M trainable parameters correspond to a 34\,MB FP32 upload per round for monolithic baselines.

The experiment uses sixteen clients across four embodiments---seven-degree-of-freedom Franka Panda and Sawyer arms and six-degree-of-freedom UR5e and Jaco arms---and four task families: tabletop pick-and-place, bin packing, stacking, and drawer manipulation. Each client holds 500 locally collected trajectories with non-overlapping object categories and distinct scene textures, producing a strongly non-IID partition. Training runs for 100 federated rounds with five local epochs per round.

All methods run on a common wireless substrate: the 3GPP Technical Report (TR)~38.901 indoor-factory path-loss model for sparse clutter and low base-station height (InF-SL) at a 7\,GHz mid-band carrier, with temporally correlated shadowing (4\,dB) and block Rayleigh fading \cite{3gpp38901}, 23\,dBm uplink transmit power, and CRC-protected packets with at most four HARQ attempts. The action stream uses a reserved 5\,MHz URLLC slice with repetition coding, while vision and language share a 100\,MHz eMBB slice under a frequency-reuse factor of one, so the canonical two-cell layout already contains co-channel interference. Rounds are synchronous with a deadline: overdue vision or language updates are deferred to their next synchronization point with staleness recorded, and overdue action updates are dropped for that round. Every method consumes identical pre-generated channel traces, making all comparisons paired. Unless otherwise stated, the average uplink signal-to-noise ratio (SNR) is 10\,dB, the jammer is off, and differential privacy is disabled (studied separately in Fig.~\ref{fig_privacy}).

We compare FedMVLA with FedAvg \cite{mcmahan2017communication}, FedProx \cite{li2020federated}, SCAFFOLD \cite{karimireddy2020scaffold}, a manipulation port of FedVLN's partial aggregation (FedVLN-P) \cite{zhou2022fedvln}, and an implementation of FedVLA's expert-driven aggregation \cite{miao2025fedvla}. All methods use the same trainable backbone, data partitions, local training budget, communication rounds, and paired channel traces while retaining their respective aggregation rules. Task success is measured over 50 evaluation rollouts per client at reported endpoints (20 at sweep points), averaged over three seeds with 95\% confidence intervals; at $N{=}128$, a 32-client subset is stratified by embodiment.

\subsection{Convergence, Scalability \& Robustness for Wireless Stacks}

\begin{figure*}[tp]
\begin{center}
\setlength{\abovecaptionskip}{+0.1cm}
\setlength{\belowcaptionskip}{-0.0cm}
\centering
  \includegraphics[width=0.85\textwidth]{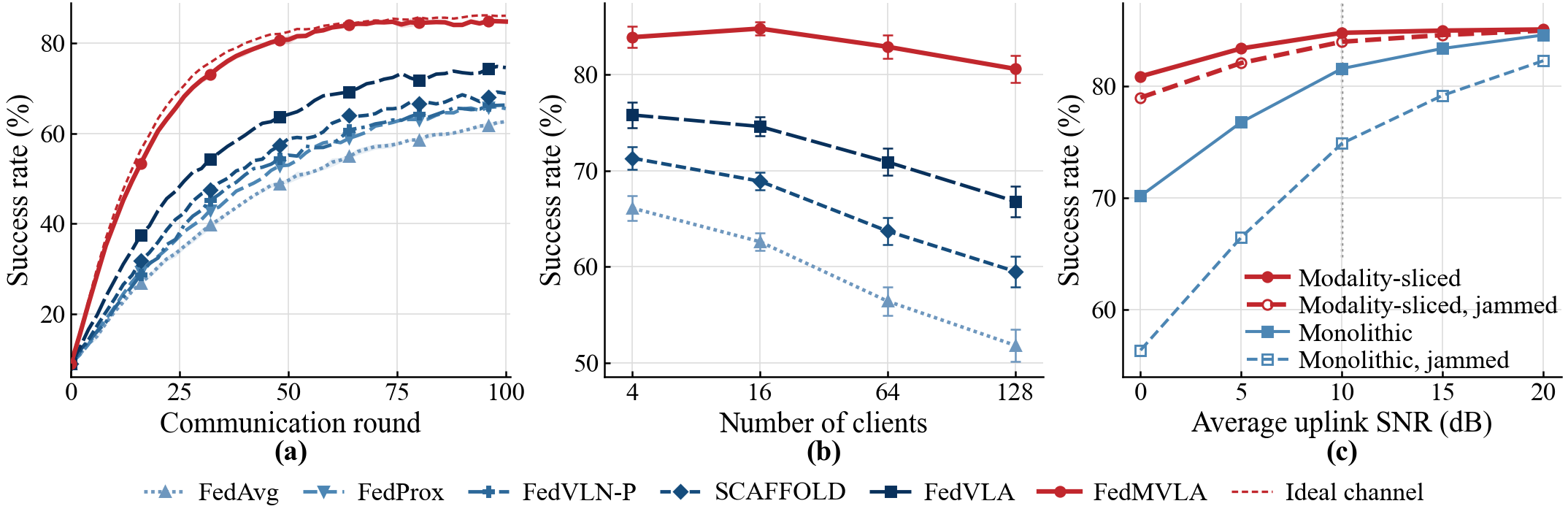}
\renewcommand\figurename{FIGURE}
\caption{\scriptsize System-level performance over the 6G wireless stack. (a) Convergence at the canonical operating point (16 clients, two cells, 10\,dB average uplink SNR); thin dashed curves are ideal-channel references. (b) Final task success rate when scaling from four clients in one cell to 128 clients across eight cells at a fixed total budget of 8,000 demonstrations. (c) Robustness to fading and pulsed jamming (duty cycle 0.3, INR 10\,dB); both transport variants use the same FedMVLA learning algorithm and differ only in transport.}
\label{fig_system}
\end{center}
\vspace{-6mm}
\end{figure*}

Fig.~\ref{fig_system}(a) compares six methods over the full wireless substrate. FedMVLA reaches 84.8\%, exceeding FedAvg (62.6\%) by 22.2 percentage points and FedVLA (74.6\%) by 10.2 percentage points. Paired ideal-channel references estimate a 1.3-percentage-point degradation for FedMVLA and 3.4 percentage points for FedAvg under the modeled wireless impairments. Orthogonal frequency planning attributes 0.4 and 1.3 percentage points, respectively, to co-channel interference, with the remaining degradation associated with fading-induced loss and staleness. At the canonical point, replacing MAFA with global averaging, MACO with iso-bitrate uniform compression, or sliced transport with monolithic transport costs 4.4, 5.3, and 3.2 percentage points; MAPA's ablation is the uniform-DP curve in Fig.~\ref{fig_privacy}(b). The measured gradient geometry is consistent with the three MAFA topologies: action-update similarity is near zero across embodiments and about 0.6 within groups, vision gradients form three scene clusters, and language gradients remain aligned above 0.8.

Fig.~\ref{fig_system}(b) fixes the total demonstration budget at 8,000 and scales from four clients in one cell to 128 clients across eight cells, with three compute tiers. Under increasing data fragmentation and inter-cell interference, FedMVLA's margin over FedAvg grows from 17.8 to 28.8 percentage points, showing more graceful degradation at scale.

Fig.~\ref{fig_system}(c) sweeps average uplink SNR from 0 to 20\,dB under a pulsed jammer (duty cycle 0.3 and interference-to-noise ratio (INR) 10\,dB). Jamming reduces modality-sliced FedMVLA by at most 1.9 percentage points; the 1.2\,MB action update uses relatively few repetition-coded URLLC transmission slots, keeping residual action-stream packet loss below $10^{-4}$ even at 0\,dB. Mean vision-update staleness remains below one synchronization period at the canonical SNR and reaches about 1.4 periods at 0\,dB. The paired monolithic transport overlaps more jammer-active slots and falls to 56.4\% in the harshest setting, quantifying the benefit of modality-sliced transport under the same learning algorithm.

\captionsetup{font={scriptsize}}

\subsection{Privacy-Utility Trade-off}

\begin{figure}[tp]
\begin{center}
\setlength{\abovecaptionskip}{+0.1cm}
\setlength{\belowcaptionskip}{-0.0cm}
\centering
  \includegraphics[width=0.85\columnwidth]{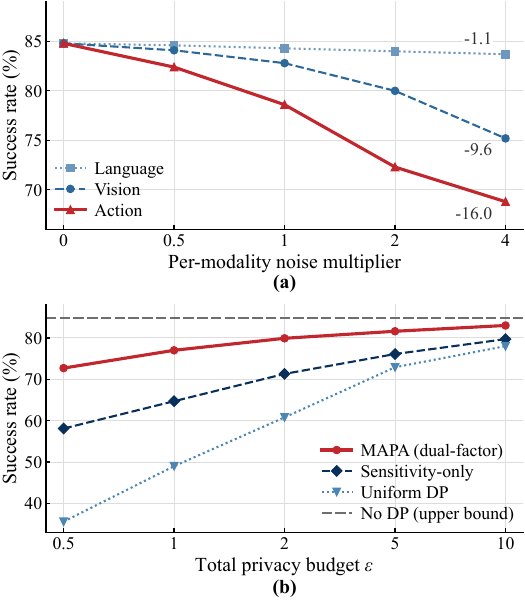}
\renewcommand\figurename{FIGURE}
\caption{\scriptsize Privacy analysis. (a) Per-modality noise tolerance: Gaussian noise of multiplier $z_m$ is injected into one modality's update only. (b) Task success rate versus total privacy budget under record-level differential privacy with RDP accounting ($\delta=10^{-5}$); the legend reports MAPA's optimized budget allocation at $\varepsilon=1$.}
\label{fig_privacy}
\end{center}
\vspace{-6mm}
\end{figure}

Fig.~\ref{fig_privacy}(a) isolates perturbation tolerance by injecting Gaussian noise into a single modality's aggregated update while leaving the others noise-free. At noise multiplier $z_m{=}4$, the language pathway loses only 1.1 percentage points, vision loses 9.6 percentage points, and action loses 16.0 percentage points. This measured asymmetry, rather than a design assumption, is what feeds MAPA's allocation optimizer, and it substantiates the claim in Section~II-B that equal-strength noise affects the pathways very unequally.

Fig.~\ref{fig_privacy}(b) reports the system-level trade-off under record-level differential privacy with RDP accounting ($\delta{=}10^{-5}$). At $\varepsilon{=}0.5$, MAPA retains 72.7\% task success versus 35.6\% under uniform allocation and 58.1\% under sensitivity-only allocation. This comparison supports combining privacy sensitivity with measured perturbation tolerance. At $\varepsilon{=}1$, the optimizer assigns 61\% of the budget to action, 31\% to vision, and 8\% to language; at relaxed budgets, MAPA approaches the 84.8\% no-privacy reference.

\captionsetup{font={scriptsize}}

\subsection{Communication Efficiency and Latency}

A compression-tolerance micro-benchmark compresses one pathway at a time from an 85.4\% all-FP32 reference. Top-1\% vision sparsification with error feedback costs 0.3 percentage points and four-bit language quantization costs 0.4 percentage points, whereas four-bit action-head quantization produces a 0.87\,mm single-step end-effector root-mean-square (RMS) error and a 10.8-percentage-point task-success drop. Together with the 16-step action chunks, these end-to-end measurements motivate retaining FP32 for the action pathway.

\begin{figure}[tp]
\begin{center}
\setlength{\abovecaptionskip}{+0.1cm}
\setlength{\belowcaptionskip}{-0.0cm}
\centering
  \includegraphics[width=0.80\columnwidth]{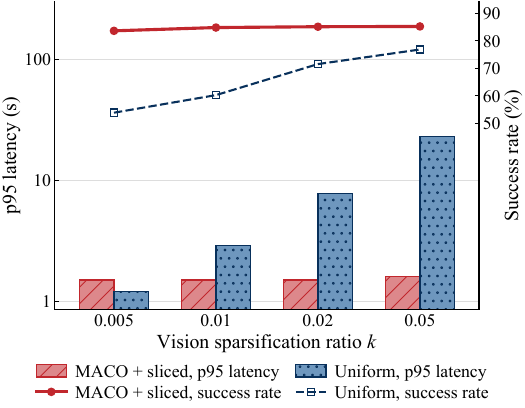}
\renewcommand\figurename{FIGURE}
\caption{\scriptsize p95 of the round-critical uplink completion time (bars, log scale) and task success rate (lines) versus the vision sparsification ratio $k$. The dotted line marks the FedAvg FP32 reference (34\,MB per round).}
\label{fig_comm}
\end{center}
\vspace{-8mm}
\end{figure}

Fig.~\ref{fig_comm} reports the 95th percentile (p95) of the round-critical uplink completion time together with task success. MACO with sliced transport remains near 1.5\,s across sparsification ratios because only the URLLC-protected action stream lies on the critical path; the uniform monolithic pipeline grows from 1.2\,s to 23\,s as $k$ increases. At $k{=}0.005$, uniform transport is faster but reduces task success to 53.8\%, whereas MACO preserves both latency and utility. FedMVLA's schedule-averaged per-client uplink payload is about 1.5\,MB per round, a 95.6\% reduction (approximately 96\%) from the 34\,MB FedAvg baseline; the latter reaches a p95 of 136\,s. Using the simulator's per-round airtime statistic, transmit energy is approximately 0.3\,J for FedMVLA and 2.7\,J for FedAvg ($E=P_{\mathrm{tx}}t_{\mathrm{air}}$).

On a held-out task family excluded from all local datasets, FedMVLA retains a 14-percentage-point advantage over FedAvg, providing a within-domain generalization check beyond the four training families. The evaluated object-handling tasks reflect elements of the household, assistive, logistics, and industrial scenarios in Fig.~\ref{fig_scenario}; mobile and navigation-centric embodiments remain an open extension.

\captionsetup{font={scriptsize}}

\section{Open Issues and Future Directions}

\subsection{Safety, Robustness \& Anti-Jamming for Action Aggregation}
Although group-wise aggregation mitigates embodiment mismatch, abnormal or malicious action updates may still cause collisions, excessive torque, unstable grasps, or workspace violations. Future systems should combine embodiment-specific constraint projection with anomaly screening before redistribution: edge servers can enforce joint, workspace, velocity, and torque limits, while the Non-RT RIC maintains safety profiles and near-real-time components perform feasibility checks. Anti-jamming protection should coordinate frequency hopping and spread-spectrum signaling at the physical layer \cite{wang2020dynamic}, RIC-based detection and slice remapping at the control layer, and staleness-aware or norm-bounded aggregation at the learning layer. Randomized slice hopping should also be evaluated against its signaling and scheduling overhead.

\subsection{O-RAN-Native Modality Orchestration}
A practical O-RAN deployment can separate long-term orchestration from near-real-time enforcement. A Non-RT RIC application (rApp) may plan privacy budgets, aggregation topology, synchronization frequency, and global modality policies and deliver them through A1. A Near-RT RIC application (xApp) may then enforce client selection, adaptive noise, slice assignment, and compression through E2. When channel quality degrades, the xApp could increase vision sparsification or postpone language synchronization while maintaining the full-precision floor and protected resources assigned to action updates. This hierarchical rApp--xApp mapping offers a concrete deployment route, although conformant interface implementation and timing evaluation remain as a series of systematic works in the future.

\subsection{Semantic, Scalable Communications for Large VLA Models}
MACO currently uses modality-specific but largely fixed compression. Future designs should adapt compression jointly to channel state, semantic importance, and task context: saliency can prioritize vision updates, quantization-aware coding can preserve language semantics, and reliability-enhanced joint source--channel coding can protect action updates. Analog over-the-air aggregation is also promising because channel noise may contribute to differential privacy below an SNR-dependent threshold \cite{liu2021privacy}, although it is outside the digital stack evaluated here. To scale larger VLA policies, LoRA, adapters, or prompt modules can reduce the trainable state, while hierarchical edge execution assigns vision, language, and action processing to tiers matching their latency and reliability needs.

\section{Conclusion}
This article investigated the structural differences among modality-specific modules within VLA models. Specifically, we proposed FedMVLA, a modality-decoupled FL framework comprising three synergistic mechanisms. MAFA mitigates cross-modality gradient interference through differentiated aggregation topologies, MAPA allocates differential privacy budgets by jointly considering modality-specific privacy sensitivity and perturbation tolerance, and MACO matches compression schemes to the precision requirements of each modality. 
A case study over a 3GPP-based wireless substrate shows that FedMVLA achieves an 84.8\% task success rate, exceeds FedAvg by 22.2 percentage points, sustains a widening margin when scaling to 128 clients across eight cells, and reduces the schedule-averaged per-client uplink model-update payload by 95.6\% (approximately 96\%), while keeping the p95 of the round-critical uplink completion time near 1.5\,s under fading, co-channel interference, and pulsed jamming. 
More broadly, modality decoupling aligns naturally with the 6G edge hierarchy, O-RAN control loops, and semantic communications, supporting privacy-preserving embodied intelligence as a native network service.

\vspace{-4mm}

\bibliographystyle{IEEEtran}
\bibliography{references.bib}

\section*{Biographies}

\vspace{-10mm}

\begin{IEEEbiographynophoto}{Zhuodong Liu} 
received the B.S. degrees in Information Systems from Beijing Jiaotong University, Beijing, China, and Rochester Institute of Technology, Rochester, USA, through a dual-degree program. He is currently pursuing the M.Phil. degree at The Hong Kong University of Science and Technology (Guangzhou), Guangzhou, China.
His research interests include federated learning, data mining, and large language models.
\end{IEEEbiographynophoto}


\begin{IEEEbiographynophoto}{Xiangyu Li} 
received the M.S. degree from Georgia Institute of Technology, Atlanta, USA, in 2023, and he is currently pursuing a Ph.D. degree at Shanghai Jiao Tong University (SJTU), Shanghai, China in the Eastern Institute of Technology, Ningbo (EIT)-SJTU Joint Ph.D. Program. His research interests include wireless communications, network security, and interdisciplinary applications of AI.
\end{IEEEbiographynophoto}


\begin{IEEEbiographynophoto}{Chunhong Yuan} 
received the B.S. degree from Ningbo University, Ningbo, China, in 2024, and he 
is currently pursuing a Master's degree at the Faculty of Control Systems and Robotics, ITMO University, St. Petersburg, Russia. His research interests include machine learning and large language models.
\end{IEEEbiographynophoto}


\begin{IEEEbiographynophoto}{Hongyang Du} (Member, IEEE) 
received the Ph.D. degree from the Nanyang Technological University, Singapore. 
He is an assistant professor at the Department of Electrical and Electronic Engineering, The University of Hong Kong, Hong Kong SAR, China. 
His research interests include edge intelligence, generative AI, and network management.
\end{IEEEbiographynophoto}


\begin{IEEEbiographynophoto}{Bodong Shang} (Member, IEEE) 
received his Ph.D. degree from Virginia Tech, Blacksburg, USA, in 2021. 
He is currently an Assistant Professor at Eastern Institute of Technology, Ningbo, Zhejiang, China. His research interests include space-air-ground-sea integrated networks, non-terrestrial networks, and space information networks.
\end{IEEEbiographynophoto}


\begin{IEEEbiographynophoto}{Qingqing Wu} (Senior Member, IEEE) 
is currently an Associate Professor with Shanghai Jiao Tong University, Shanghai, China. 
He has co-authored more than 100 IEEE journal articles with more than 40 ESI highly cited papers, which have received more than 50,000 Google citations. 
He is the Founding Chair of IEEE Communications Society Young Professional committee in Asia–Pacific Region and the Chair of IEEE VTS Drone Committee.
\end{IEEEbiographynophoto}


\begin{IEEEbiographynophoto}{Tony Q. S. Quek} (Fellow, IEEE) 
is currently the Cheng Tsang Man Chair Professor with Singapore University of Technology and Design and a ST Engineering Distinguished Professor. 
His current research topics include wireless communications and networking, network intelligence, non-terrestrial networks, open radio access network, and 6G. 
He is a fellow of WWRF and the Academy of Engineering Singapore.
\end{IEEEbiographynophoto}


\begin{IEEEbiographynophoto}{Mohsen Guizani} (Fellow, IEEE) 
is currently a Professor and an Associate Provost with the Mohamed Bin Zayed University of Artificial Intelligence, Abu Dhabi, United Arab Emirates. His research interests include applied machine learning, smart city, wireless communications/networking, cloud computing, and security. He is also serving on the editorial boards for many IEEE Transactions and magazines.
\end{IEEEbiographynophoto}

\end{document}